\documentclass{article}

\usepackage[final,nonatbib]{neurips_2020}

\usepackage[utf8]{inputenc} %
\usepackage[T1]{fontenc}    %
\usepackage{hyperref}       %
\usepackage{url}            %
\usepackage{booktabs}       %
\usepackage{amsfonts}       %
\usepackage{nicefrac}       %
\usepackage{microtype}      %
\usepackage{graphicx}
\usepackage{wrapfig}
\usepackage{enumitem}
\usepackage{amsmath}
\usepackage{bbm, dsfont}
\usepackage{caption}
\usepackage{subcaption}
\usepackage{tikz}
\usepackage{pgfplots}
\usepackage{float}
\usepackage{graphicx}
\usepackage{caption}
\usepackage{textcomp}
\usepackage{pdfpages}
\usepackage{enumitem}
\setlist{leftmargin=5.5mm}
\usepackage[ruled]{algorithm2e}

\usepackage{multirow}
\usepackage{xcolor}
\usepackage{xspace}
\usepackage{units}

\definecolor{myred}{rgb}{0.8,0,0}
\definecolor{mygreen}{rgb}{0,0.6,0}
\definecolor{myblue}{rgb}{0,0,0.7}

\newcommand{\rosetta}{{\sc ROSETTA}\xspace}
\newcommand{\sarscovv}{{\sc SARS-CoV-2}\xspace}
\newcommand{\sarscov}{{\sc SARS-CoV-2}\xspace}
\newcommand{\ppo}{{\sc ppo}\xspace}

\newcommand{\sppo}{{\sc sppo}\xspace}

\title{Designing a Prospective COVID-19 Therapeutic with Reinforcement Learning}

\author{Marcin J. Skwark\thanks{correspondence: m.skwark@instadeep.com}\\
    InstaDeep
    \And 
    Nicolás López Carranza\\
    InstaDeep
    \And 
    Thomas Pierrot\\
    InstaDeep
    \And 
    Joe Phillips\\
    InstaDeep
    \And 
    Slim Said\\
    InstaDeep
    \And 
    Alexandre Laterre\\
    InstaDeep
    \And 
    Amine Kerkeni\\
    InstaDeep
    \AND
    Uğur Şahin\\
    BioNTech
    \And
    Karim Beguir\\
    InstaDeep
}

\begin{document}

\maketitle

\begin{abstract}

The \sarscovv pandemic has created a global race for a cure. One approach focuses on designing a novel variant of the human angiotensin-converting enzyme 2 (ACE2) that binds more tightly to the \sarscovv spike protein and diverts it from human cells. %
Here we formulate a novel protein design framework as a reinforcement learning problem. We generate new designs efficiently through the combination of a fast, biologically-grounded reward function and sequential action-space formulation. The use of Policy Gradients reduces the compute budget needed to reach consistent, high-quality designs by at least an order of magnitude compared to standard methods. Complexes designed by this method have been validated by molecular dynamics simulations, confirming their increased stability. This suggests that combining leading protein design methods with modern deep reinforcement learning is a viable path for discovering a Covid-19 cure and may accelerate design of peptide-based therapeutics for other diseases.

\end{abstract}

\section{Introduction}
\vspace{-1em}

The ongoing worldwide Covid-19 pandemic is compelling researchers to explore multiple routes to develop a cure. While the creation of small molecule drugs remains a promising path \cite{ton2020rapid,zhou2020network,wang2020fast}, another possibility is the design of a novel peptide that would bind to and neutralize the SARS-CoV-2 spike protein \cite{huang2020computational, han2020computational}. To replicate, the virus needs to penetrate the cell by hijacking its active transport processes. It does so through a specific set of interactions with the cellular receptor. For SARS-CoV and SARS-CoV-2, it is the spike protein that mediates the cell infection through the binding of its receptor-binding domain (RBD) to the angiotensin converting enzyme 2 (ACE2) on the surface of the cell \cite{wrapp2020cryo}.
\sarscovv only recently crossed the species barrier into humans and developed an affinity towards ACE2, meaning its binding is likely to be suboptimal. A new binder could be engineered to neutralize the SARS-CoV-2 virus by inhibiting its attachment process to the native ACE2. To do so, it would need to be more specific for \sarscovv, i.e. have a higher binding affinity than ACE2, while simultaneously remaining sufficiently similar to the native variant in order to be safe for humans. The resulting binder would keep \sarscovv outside the human cells, making it more visible for recognition by the immune system. The interface between ACE2 and the \sarscovv spike protein consists of 34 amino acids, with 20 residues possible at each position. Designing the interface of a novel variant of ACE2 therefore results in a search space over amino acids in the order of $20^{34}$, or equivalently, $10^{45}$.

To explore this research direction, we have developed a novel protocol, which given a protein complex structure, optimizes the binding interface. This is achieved through the combination of computational structural biology and deep reinforcement learning (RL).  The protocol maintains the stability of the interacting partners, and optimizes binding while minimizing the number of redesigned positions, leading to designs that are more stable, and due to minimal, unobtrusive quality of the design, less likely to either misfold, or elicit an immune reaction \emph{in vivo}. While it is not realistic to directly estimate the binding affinity between interacting partners, one can relatively well approximate the free energy of the system. We leverage the industry standard methodology implemented in the ROSETTA suite \cite{alford2017rosetta, chaudhury2010pyrosetta}
to evaluate quality of binding through a combination of standard terms present in the ROSETTA energy function.

\section{Related work}
\vspace{-1em}

To the best of our knowledge, the only work applying Reinforcement Learning to protein design can be found in \cite{dynappoprotein}. The authors propose an RL formulation, but postulate that the evaluation cost of the scoring function hinders its applicability. Consequently, to improve the sample efficiency, they developed a model-based approach. The scoring function is approximated using an ensemble of models of various nature and capacities. This reduces the estimation errors, that could have been exploited by the agent. We differ from the resulting method (DynaPPO) not only in terms of the problem formulation -- more expressive action space and initial state space, which increase sample efficiency -- but also in terms of the scoring function, which balances accuracy and speed, obviating the need for approximation with ensemble models.

\section{Problem Formulation}
\label{sec:design}

\begin{figure*}[h!]
    \centering
    \includegraphics[width=0.8\linewidth]{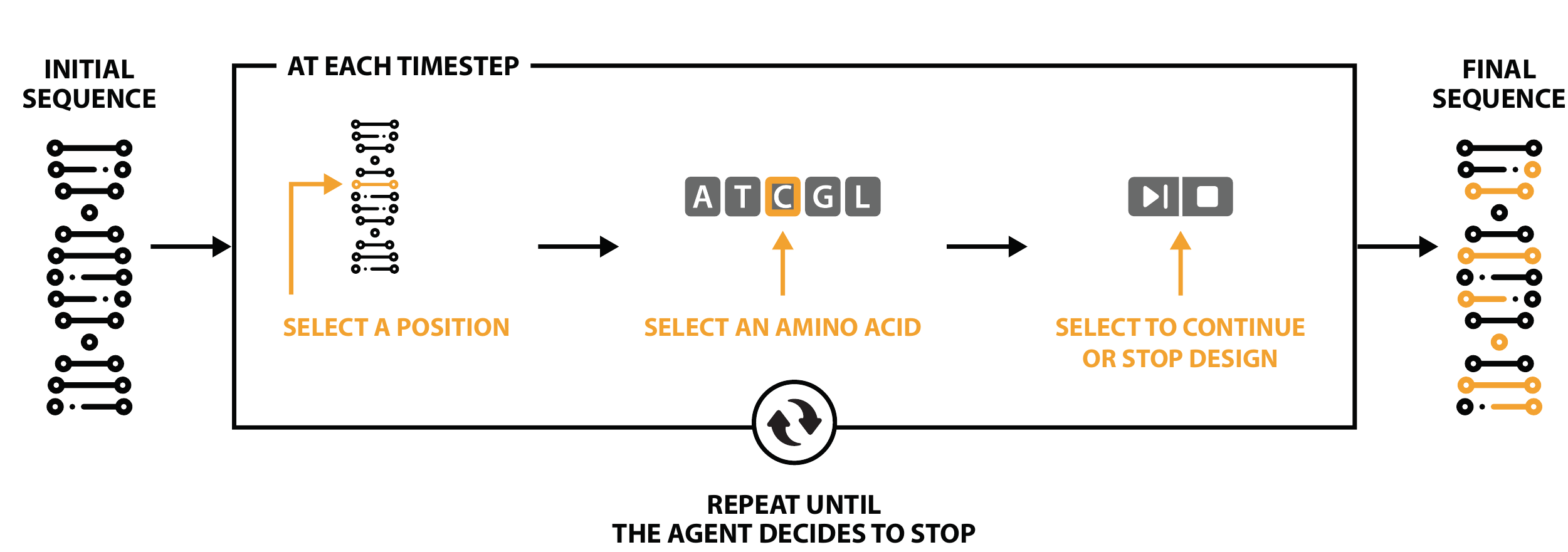}%
    \caption{\small Protein design expressed as a Markovian Decision Process. Figure shows the \emph{iterative} problem formulation, where the first state is initialized with a sequence to be mutated by the RL agent. Here we depict the sequential action space used by \sppo, where the policy is factorized over the decisions of which position to select, what to change it to, and when to stop mutating.}
    \label{fig:action_space}
\end{figure*}

We formulate the design of high affinity binder proteins as a Markov Decision Process (MDP) $\left( \mathcal{S}, \mathcal{A}, \mathcal{T}, R, \gamma \right)$ where $\mathcal{S}$ is the state space, $\mathcal{A}$ the action space, $\mathcal{T}: \mathcal{S} \times \mathcal{A} \rightarrow \mathcal{S}$ the transition function, $R$ the reward signal and $\gamma$ a discount factor. One episode corresponds to the design of a protein. An environment state represents a protein, i.e. a sequence of amino acids $s = \left( s^1, s^2, \dots, s^{n} \right)$, where $s^i$ refers to the residue at position $i$ and $n$ is the length of the complete sequence. Each amino acid is chosen among a set of $m=20$ possible proteinogenic types. A possible formulation (e.g. \cite{dynappoprotein}), is to initialize each episode with a blank sequence and then generate each amino acid progressively. In this case, the agent chooses $n$ times between $m$ actions during each episode. At the end of an episode, the final protein obtained is scored. The resulting reward signal is sparse, i.e non-zero only at the end of an episode. The reward is normalized to the interval $R\in(-1.0, 1.0]$, with ACE2 set as a baseline at $R_{\text{ACE2}}=0$. In our work, we refer to this formulation as the \emph{generative} approach.

In this study, we propose another formulation: the \emph{iterative} approach. In this setting, the agent begins each episode with an initial sequence. The agent then \emph{mutates} the sequence by selecting positions to change, as well as the new residues to change them to, in such a way that there is always a valid sequence to be scored. As a result, the mutation process can be stopped at any moment, potentially by the agent itself (see figure \ref{fig:action_space}). The choice of initial sequence can be arbitrary, though we choose either the native human ACE2 protein sequence, or randomly perturbed versions thereof. Prior initialization conditioned on ACE2 enables the agent to prioritize exploration around the subset of states that represent close mutations of the native ACE2 protein sequence, which is the only protein known to bind to \sarscovv with certainty.

Contrary to \cite{dynappoprotein}, our target function is not learned, but computed directly from simulation. Protein-protein interactions are driven by a large set of intertwined interactions, which are practically impossible to estimate from protein sequence space alone. Our scoring function therefore estimates the change of Gibbs free energy %
resulting from the action chosen by the agent. This is computed through the combination of knowledge-based statistical potential \cite{alford2017rosetta}, and prior knowledge of experimentally observed protein complex conformations derived from Protein Data Bank \cite{berman2000protein}, as well as binding target variability \cite{hatcher2017virus}. The score reflects the binding-oriented problem formulation, which expressly does not capture the entire complexity of the physiology of virus adhesion to the cell.

\section{Methods}\label{sec:rl}

\paragraph{Reinforcement Learning Algorithms}
For both formulations of the protein design problem, we make use of policy gradient methods \cite{reinforce} to directly optimize a stochastic policy parametrized by vector $\theta$, $\pi_{\theta}: \mathcal{S} \rightarrow \mathcal{D}_{\mathcal{A}}$ where $\mathcal{D}_{\mathcal{A}}$ is the space of distributions over the action space $\mathcal{A}$. The iterative formulation defined in section \ref{sec:design}, also allows the problem to be solved using sequential action spaces, i.e. the agent must select not one action, but a sequence of $p$ actions. At each step, the agent must first choose a position in the sequence, then select an amino acid to place, given the chosen position, and finally, it must decide whether to take the action to stop the design. When the agent chooses to stop, the episode terminates and the resulting sequence is scored.
At each time step, the three actions are selected autoregressively, each one depending on the choices of the previous ones. The factorization of the action space helps to reduce the number of parameters of the policy network. It also improves decision interpretability and eases the learning by splitting one potentially extremely large distribution into several smaller conditional distributions. Empirically, it displays improved exploration capacities as well.

We implement the policy gradient method using Proximal Policy Optimization (PPO) \cite{ppo}, and extend PPO to the sequential setting for the sequential action space configurations. We call the resulting algorithm \emph{Sequential PPO} (SPPO). We show that SPPO learns in a few time-steps to start from random perturbations of the ACE2 sequence and make only a few changes in its amino acids to obtain good scores.

\paragraph{Neural network architecture}
Attention-based model architectures, such as Transformers \cite{vaswani2017attention} have recently achieved state-of-the-art results in NLP across various tasks \cite{bert, xlnet, universal_trans}.
In our work, we leverage the ability of transformer models to process sequential information and extract long-range dependencies. The transformer backbone is followed by the appropriate heads, e.g. Policy and Value heads for (S)PPO, as shown in figure \ref{fig:rl_train}. A complete description of the network architecture alongside its hyper-parameters and the input specifications can be found in section \ref{appendix:neural} of the supplementary materials.

\section{Experimental results}
\begin{figure}[bt]
\begin{minipage}{\textwidth} 
\begin{figure}[H]
    \centering
    \includegraphics[height=0.9cm]{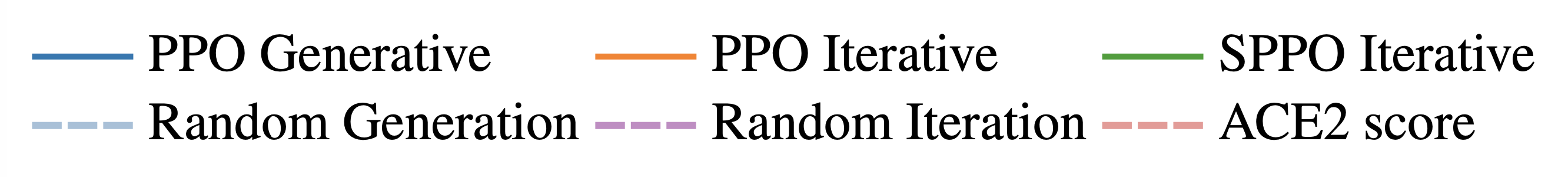}
\end{figure}
\end{minipage}
\begin{minipage}{.47\textwidth} 
\begin{figure}[H]
    \centering
    \newlength\fheight
    \newlength\fwidth
    \setlength\fheight{4.75cm}
    \setlength\fwidth{\linewidth}
    \input{figures/average}
\end{figure}
\end{minipage}
\hfill
\begin{minipage}{.47\textwidth}
\vspace{-0.4cm}
\begin{figure}[H]
    \centering
    \setlength\fheight{4.75cm}
    \setlength\fwidth{\linewidth}
    \input{figures/maximum}
\end{figure}
\end{minipage}
\vspace*{-5mm}
\caption{\small Average and maximum episode return, demonstrating the performance of \ppo and \sppo reinforcement learning algorithms on the generative and iterative formulations of the protein design MDP. Reward, $R\in{(-1.0, 1.0]}$. Native ACE2 is benchmarked with a score of $R_{\text{ACE2}}=0$. In all cases, RL agents learn to design superior binders than the native ACE2 and outperform baseline agents based on sampling.}
\label{fig:tensorboard}
\end{figure}

On both formulations of the protein design problem, the RL agents exceed the performance of a set of benchmark methods based on simulated annealing, Monte-Carlo sampling, and random sampling. These results are shown in Figure \ref{fig:tensorboard} and in Appendix \ref{appendix:rl_plots}. For the iterative formulation of the problem, we note that the process of mutating sequences initialized from native ACE2 introduces a source of stability. This formulation reduces the variance in the mean episodic reward by two orders of magnitude, compared to the generative formulation of the problem.
While both PPO and SPPO achieve comparable scores, SPPO yields candidate designs above the threshold of native ACE2 in a much more sample efficient manner. The factorization of the action space into sequential conditional policies enables more informative gradient updates during training, compared to those received in the joint action space configuration implemented by PPO. Indeed, autoregressive policies can offer more consistent exploration in RL tasks, especially when there are sparse rewards~\cite{Korenkevych2018autoreg}. This results in a 5-10x reduction in the number of steps required to surpass the native ACE2 score.

\section{Design Validation}
\label{sec:validation}

We compared the best-scoring designs of our method to designs previously reported in \cite{huang2020computational, han2020computational}. We applied the industry-standard \rosetta \texttt{remodel} \cite{huang2011rosettaremodel} method to the same reference structures, and optimize the same positions as in this work. To ensure a conservative comparison with our results, we devoted ten times more CPU hours to the Rosetta designs than recommended by \cite{heavey_2007}, starting the designs from multiple, alternative conformations, including those derived from experimental structures, those used internally in our solution, and their variants obtained by conformational perturbation and minimization. While this protocol diverges from state-of-art \rosetta approaches, it allows us to estimate the computational effort needed to tackle such a problem with no external expert input. Therefore we claim that, given the typical computation time recommended for such tasks, reinforcement learning delivers better results with greater diversity than traditional methods~(see Table \ref{table:speed}).

\paragraph{Knowledge-based assessment}

For each design candidate (either ours, published, or generated by \rosetta), we construct a pool of structural models. Each model is subject to the \rosetta \texttt{InterfaceAnalyser} protocol~\cite{stranges2013comparison}.  This shows that our designs both bind ACE2~(see annex  \ref{tab:methods_templates_comp}), and retain sufficient similarity to human ACE2 receptor, to evade recognition by immune system, and to make them plausible as \emph{drop-in} therapeutics.
In terms of the Gibbs free energy ($\delta G$) of binding, our designs, proposed in an entirely automated manner, prove to be \emph{substantially more competent binders than human ACE2}. We note, that it is our method that proposed \emph{the best performing design} in terms of $\delta G$ in the entire benchmark. The design proposed in \cite{huang2020computational} outperforms us in terms of EvoEF2 and RW+ metrics, however these are the metrics authors have been explicitly optimizing, while we have not been optimizing for binding explicitly.

\paragraph{Molecular dynamics simulations}
\label{sec:md}
The analysis above relies on a knowledge-based potential and, as such, involves theoretically optimized structures, which can lead to overfitting. To assess the binding performance of newly designed proteins, we run a series of unrestrained, molecular dynamics simulations using the industry standard, GPU-aided GROMACS 2018.4 \cite{abraham2015gromacs,pall2014tackling,pronk2013gromacs} software package. 
In a 50ns simulation (on average over 5650 core-hours per simulation) the native human ACE2 does not bind particularly strongly to \sarscovv RBD. After 15ns of simulation time, partners drift apart, to fully disconnect by 20 ns (see Figure~\ref{fig:mdVisualisation}). We see evidence of complex reformation by the end of the trajectory, as partners come back together. Our designs do not exhibit such a behavior and are more stable, as measured by the trajectory RMSD to the bound conformation. Even if started from suboptimally bound pose, partners with newly designed interfaces quickly form an transitory complex and remain stable throughout the simulation.  The analyses of molecular dynamics trajectories indicate that our designed proteins bind \sarscovv RBD in a stronger, more affine manner. They result in much more stable complexes as well. This achieves the goal of tight binding for coronavirus neutralization.

\section{Conclusion}

In this paper we introduce a novel framework for protein design by combining reinforcement learning with computational structural biology and apply it to develop a candidate cure for Covid-19. First, we build a fast reward scoring function that leverages insights from co-crystal structures of the \sarscovv spike protein complexed with the human ACE2. Second, we experiment with several RL algorithms and demonstrate convergence. In particular, we design a sequential version of the Proximal Policy Optimization algorithm for the protein design task, which we name Sequential PPO (SPPO). SPPO displays improved sample efficiency compared to standard PPO and quickly obtains promising ACE2-based protein binders. Third, we subject our best designs to a range of evaluations and report favourable \emph{in silico} knowledge-based metrics, including for metrics not used in our scoring protocol. Our designs are competitive with those published by well-established groups in computational structural biology. Finally, full scale molecular dynamics simulations confirm that our candidate therapeutics are more stable than the native human ACE2 and bind tightly to the SARS-COV-2 RBD, suggesting our candidate cures are likely to perform their intended task \emph{in vivo}.

{\small
\bibliographystyle{plain}
\bibliography{biblio}
}

\clearpage
\newpage

\appendix
\section*{Supplementary material}
We give more details about the algorithms we use in our methods, about the baselines we compare to and about our experimental details in the following sections\footnote{Additional videos and illustrations available at \url{https://sites.google.com/view/covid-rl-design}.}.

\section{Reinforcement Learning Methods}
\subsection{Neural network architecture}
\label{appendix:neural}
\begin{figure*}[hpt]
\centering
 \includegraphics[width=0.8\linewidth]{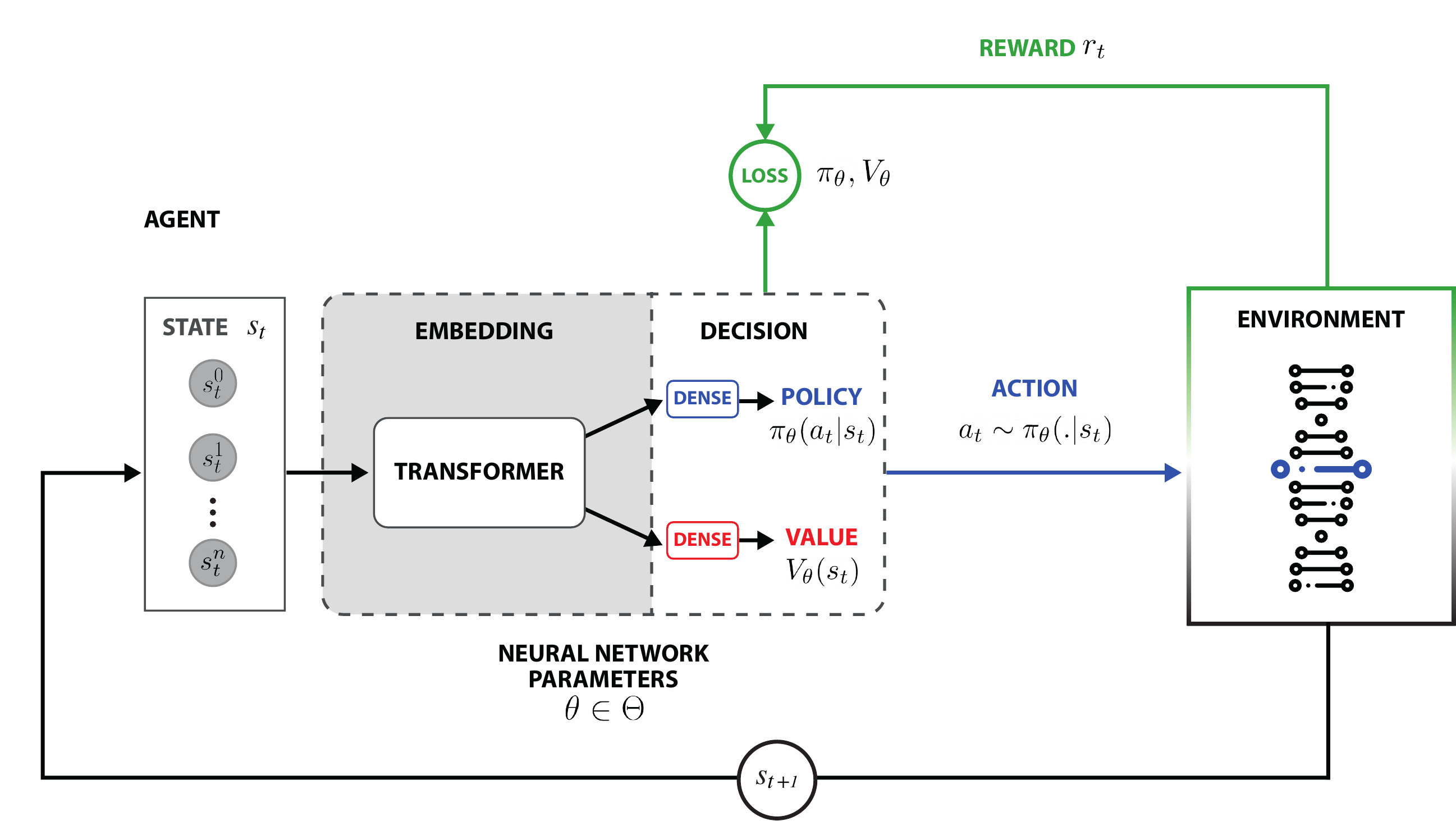}%
\caption{Protein sequence design schema of the reinforcement learning agent and environment.} %
\label{fig:rl_train}
\end{figure*}

The input of the neural network is a $n\times f$ sequence, consisting of $f=10$ features for each amino acid\footnote{When the generative formulation is used, the missing sequence elements are padded with zeros.}. Those features form the amino acid vocabulary and correspond to the Atchley factors \cite{Atchley6395}, a small set of highly interpretable numeric patterns of amino acid variability. The sequence containing $n$ amino acids is embedded as a sequence of $n$ corresponding vectors in $\mathbb{R}^d$, where $d=512$. Each vector is the sum of a learned token embedding and a sinusoidal positional embedding as in the original Transformer architecture \cite{vaswani2017attention}. The embedded sequence is then processed by a GTrXL model \cite{gtrxl}, comprising 5 blocks, each composed of a self-attention operation followed by a position-wise multi-layer network. Each block preserves the embedding dimension of each sequence element. The associated hyper-parameters can be found in table \ref{table:hyperparameters_table} and as well as those defining the appropriate policy and value heads for \ppo and \sppo. A schema is shown in figure \ref{fig:rl_train} with the policy and value heads for \ppo.

\subsection{Reward function}
\label{appendix:reward}

A key aspect of this work is that the reward function can directly output scores from \rosetta simulations, relatively quickly: in less than 1 minute and without the need of a proxy. Each design is evaluated in a series of biologically plausible scenarios, which reflect the current state of knowledge of \sarscovv RBD-ACE2 structures \cite{wang2020structural}, \cite{lan2020structure}, \cite{yan2020structural}, \cite{shang2020structural} and prior knowledge on the biochemically relevant interactions between partners. This comprises effective use of diverse structural information, leveraging experimental information on evolutionary constraints on the diversity of the partners, as well as the data on the genetic diversity of the population.  The function is designed to be progressively more accurate and has a built-in \emph{early stop} criteria. Therefore, designs that are highly unlikely to result in high scores, are evaluated in a faster and more cursory manner, while the bulk of computational resources are devoted to fine-grained evaluation of prospectively successful designs. 

The reward function is fully deterministic and provided the same score given the same input. This permitted for in-memory reward caching. For the memory cache, we use Redis, an in-memory key-value database. Depending on the RL algorithm, this allowed speedups of up to 3 times.

We ran our simulations in machines ranging from 96 cores to 224 cores. In these conditions, an average distributed reward calculation would take around 1 minute for our \rosetta configuration. If the score was previously calculated the cache would provide the reward in a few milliseconds.

\subsection{Hyperparameters}

\begin{table}[ht!]
\centering
{\small
\begin{tabular}{|l|l|r|}
\hline
\textbf{Notation} & \textbf{Description} & \textbf{Value} \\
\hline
\noalign{\vskip 2mm}
\hline
\multicolumn{3}{|l|}{\textbf{PPO, SPPO}}\\
\hline
$\gamma$ & discount factor & 0.99\\
{\it train batch size} & train batch size & 4000\\
{\it sgd batch size} & sgd batch size & 128\\
$lr$ & learning rate & $0.0001$\\
{\it clip param} & PPO clip param & 0.3\\
\hline
\noalign{\vskip 2mm}
\hline
\multicolumn{3}{|l|}{\textbf{Shared Neural Network}}\\
\hline
d & Embedding dimension & 512\\
blocks & Number of stacked \textit{GTrXL} blocks & 5\\
hidden units & Hidden units of the Position-wise MLP & 512\\
activation & Non-linearity of the Position-wise MLP & ReLU\\
\hline
\noalign{\vskip 2mm}
\hline
\multicolumn{3}{|l|}{\textbf{PPO Model}}\\
\hline
actor head & hidden layers size & 256/256\\
critic head & hidden layers size & 256/256\\
\hline
\noalign{\vskip 2mm}
\hline
\multicolumn{3}{|l|}{\textbf{SPPO Model}}\\
\hline
context dim & representation of observations passed to autoregressive policies & 32\\
shared layers & hidden layers size & 256\\
autoregressive layers & hidden layers size & 256\\
critic head & hidden layers size & 256/256\\
\hline
\noalign{\vskip 2mm}
\noalign{\vskip 2mm}
\end{tabular}
    }
\vspace{1mm}
\caption{Hyperparameters Table.}
\label{table:hyperparameters_table}
\end{table}

\subsection{Discussion of Reinforcement Learning Results}
\label{appendix:rl_plots}

\subsubsection*{Results: Reinforcement Learning vs. Baseline Methods}

\begin{table}[ht!]
    \centering
    \begin{tabular}{l|c|ccccccc}
    \hline \\
     Algorithm
     & Env.
     & $R_{\text{max}}$
     & $R_{\text{avg}}$
     & \multicolumn{1}{p{1.5cm}}{\centering Steps to \\  $R_{\text{max}} (10^3)$}
    & \multicolumn{1}{p{1.5cm}}{\centering $R_{\text{max}}$ > $0$ \\ ($10^3$ steps)}
     & \multicolumn{1}{p{1.5cm}}{\centering $R_{\text{avg}}$ > $0$ \\ ($10^3$ steps)}
     \\
    \hline
    PPO & gen & 0.069 & 0.063 & 398 & 94 & 202.5\\ %
    \hline
    PPO & iter & \textbf{0.071} & \textbf{0.067} & 317 & 18 & 72.5\\ %
    \hline
    SPPO & iter & 0.067 & 0.063 & \textbf{227} & \textbf{9.5} & \textbf{46.7}\\ %
    \hline
    Sim. Annealing & iter & 0.044 & - & - & - & -\\ %
    \hline 
    Rosetta MC & iter & 0.056 & 0.0094 & - & - & -\\ %
    \hline 
    Random & gen & -0.020 & -0.12 & - & - & -\\ %
    \hline
    \end{tabular}
    \vspace{0.3em}
    \caption{Summary statistics for RL algorithms compared to benchmark methods. Rewards displayed for $R\in{(-1.0, 1.0]}$. Native ACE2 benchmark score is $0$: $R_{\text{ACE2}} = 0$. All supplied RL figures are averaged over 3 random seeds. Simulated annealing performed with temperature 1\% and 10\%, best results shown here (1\%).}
    \label{tab:rl_rewards}
\end{table}
We note in table \ref{tab:rl_rewards}, the superior performance of the RL methods over the benchmark methods, as measured by $R_{\text{max}}$ and $R_{\text{avg}}$.
The extension of \ppo-generative to \sppo-iterative reduces the number of steps required to generate candidate protein sequences that outperform the native ACE2 at binding to SAR-CoV-2.

\section{Additional experimental details}
\label{app:add_exp_results}

\subsection{Results per assessment structure}
\label{appendix:results}
Each of the tables represents the same designs, in the same order, scored according to diverse criteria, scored per each assessment target, and in terms of a mean of all the columns.

\subsubsection*{Relative binding affinity: $\delta G$ per 100 \AA$^2$}
\label{tab:methods_templates_comp}
\begin{tabular}{lrrrrrrrrr|r}
\toprule
Method & $t_1$ & $t_2$ & $t_3$ & $t_4$ & $t_5$ & $t_6$ & $t_7$ & $t_8$ & $t_9$ & mean \\
\midrule
            PPO & -4.7 & -4.1 & -7.4 & -4.2 & -4.4 & -6.9 & -3.8 & -4.4 & -4.6 & -5.0 \\
            PPO & -4.6 & -4.0 & -4.9 & -4.7 & -8.4 & -6.7 & -3.5 & -3.5 & -4.8 & -5.0 \\
            PPO & -4.5 & -3.7 & -6.4 & -5.7 & -4.2 & -6.7 & -4.1 & -3.5 & -4.5 & -4.8 \\
                 \hline
          SPPO & -4.1 & -3.6 & -4.8 & -4.6 & -4.9 & -4.6 & -4.9 & -3.6 & -5.0 & -4.5 \\
          SPPO & -4.5 & -3.9 & -4.9 & -4.5 & -4.4 & -4.5 & -3.3 & -3.6 & -4.6 & -4.2 \\
          SPPO & -4.1 & -3.6 & -4.9 & -4.3 & -4.1 & -6.0 & -3.4 & -3.4 & -4.5 & -4.3 \\
                \hline
 Rosetta design & -4.5 & -4.0 & -6.9 & -4.8 & -4.3 & -4.6 & -3.8 & -3.7 & -5.1 & -4.6 \\
 Rosetta design & -4.5 & -4.1 & -4.3 & -6.3 & -3.4 & -6.6 & -3.9 & -4.0 & -3.9 & -4.5 \\
 Rosetta design & -4.2 & -3.8 & -5.0 & -4.3 & -6.7 & -4.1 & -4.1 & -3.9 & -4.5 & -4.5 \\
      \hline
         Native & -4.7 & -3.9 & -5.3 & -4.6 & -5.1 & -4.6 & -3.4 & -3.2 & -5.1 & -4.4 \\
              \hline
      Han\&Kral & -4.0 & -4.2 & -4.0 & -6.3 & -4.4 & -4.9 & -6.9 & -4.0 & -4.1 & -4.7 \\
          Huang & -4.0 & -3.5 & -4.1 & -3.9 & -4.5 & -6.4 & -4.1 & -4.1 & -6.6 & -4.6 \\

\bottomrule
\end{tabular}

\begin{figure}[hb!]
\centering
\includegraphics[width=.97\linewidth]{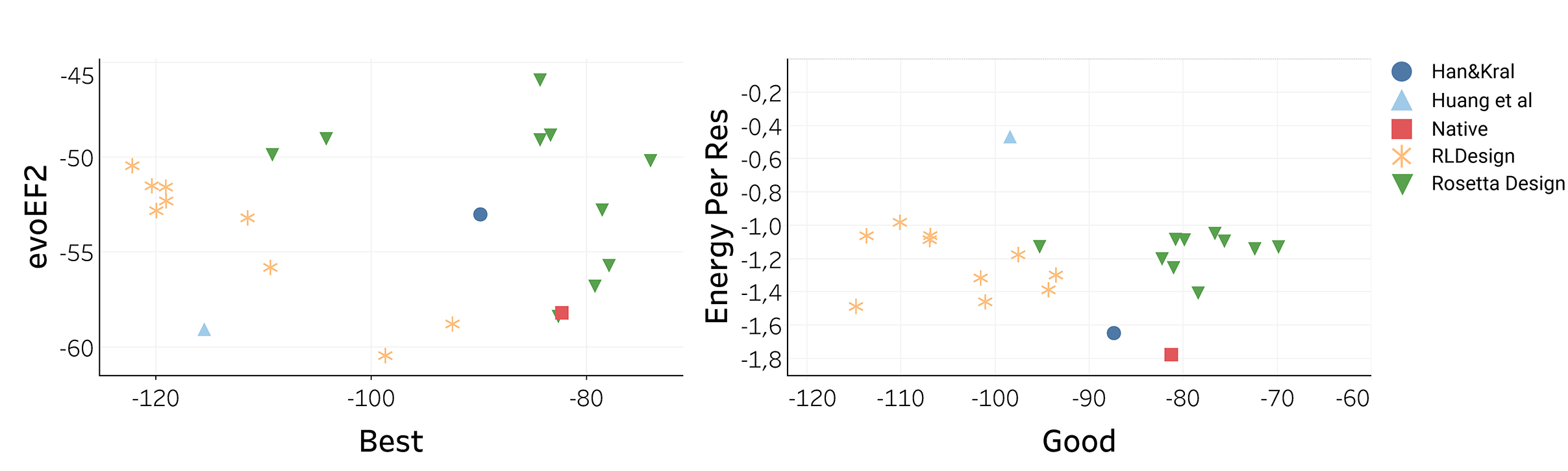}
\caption{Objective quality comparison between discussed methods. Better designs have lower scores (i.e are in the bottom left). Measures of binding energy: $\delta G$ for the best design (\emph{Best}) and mean of top 20\% designs in terms of our selection protocol (\emph{Good}) indicate high affinity.  \emph{Energy per residue} \cite{stranges2013comparison} is an estimate of the overall quality of the model.  The~\emph{evoEF2} metric is an independent estimate of binding energy~\cite{huang2020evoef2}.
With respect to every metric (except for \emph{energy per residue}, which can be alleviated by further minimization), our designs outperform the other methods, despite not optimizing any of them explicitly.}
\label{fig:bigplot}
\end{figure}

\begin{table}[hb!]
    \begin{center}
    \begin{small}
    \begin{sc}
    \begin{tabular}{c|c| c c c c}
    \toprule
    & core-hours & unique & good & designs & good designs  \\
    & (total) & designs & designs & per core-hours & per core-hours \\
    \midrule
    RLDesign (ours) &  ~1,500  & 1604  & 54 &  1 & 0.037\\
    RosettaDesign & ~40,000 & 1189 & 11 & 0.03 & 0.00027\\ 
    \bottomrule
    \end{tabular}
    \end{sc}
    \end{small}
    \end{center}
        \vspace{0.3em}
    \caption{Compute efficiency. Within a much shorter execution time, our solution delivers 50\% more unique designs than \rosetta and proposes five fold more designs that prove to be better binders than ACE2.}
\label{table:speed}
\end{table}

\subsubsection*{Absolute binding affinity: $\delta G$}

\begin{tabular}{lrrrrrrrrr|r}
\toprule
Method & $t_1$ & $t_2$ & $t_3$ & $t_4$ & $t_5$ & $t_6$ & $t_7$ & $t_8$ & $t_9$ & mean \\
\midrule

            PPO & -81.5 & -58.3 & -120.4 &  -75.0 &  -73.9 & -118.4 & -40.5 & -56.7 &  -83.6 & -78.7 \\
            PPO & -71.1 & -60.4 &  -86.0 &  -70.7 & -117.1 & -111.6 & -45.4 & -52.8 &  -80.2 & -77.3 \\       
            PPO & -75.1 & -64.4 & -113.2 &  -97.1 &  -70.7 & -109.4 & -50.4 & -58.6 &  -79.0 & -79.8 \\
                   \hline
           SPPO & -67.7 & -61.8 &  -85.8 &  -77.5 &  -91.9 &  -79.7 & -42.4 & -46.7 &  -89.0 & -71.4 \\
           SPPO & -73.7 & -62.4 &  -89.5 &  -75.6 &  -74.9 &  -74.3 & -49.8 & -50.4 &  -80.5 & -70.1 \\
           SPPO & -69.4 & -55.4 &  -85.6 &  -73.9 &  -63.6 &  -98.7 & -42.8 & -44.2 &  -81.3 & -68.3 \\
                  \hline
 Rosetta design & -77.5 & -64.7 & -111.3 &  -80.7 &  -74.5 &  -79.2 & -64.0 & -53.2 &  -88.7 & -77.1 \\
 Rosetta design & -74.2 & -65.3 &  -73.8 & -105.2 &  -59.7 & -109.2 & -49.5 & -43.0 &  -70.4 & -72.3 \\
 Rosetta design & -69.6 & -59.0 &  -88.4 &  -70.3 & -104.6 &  -68.3 & -53.1 & -52.7 &  -78.3 & -71.6 \\
        \hline
         Native & -79.0 & -63.9 &  -88.5 &  -78.1 &  -90.0 &  -76.3 & -41.8 & -45.7 &  -89.3 & -72.5 \\
                \hline
       Han\&Kral & -67.6 & -57.7 &  -66.1 &  -89.9 &  -76.1 &  -86.1 & -80.8 & -46.7 &  -75.1 & -71.8 \\
              \hline
          Huang & -62.9 & -57.8 &  -75.1 &  -58.4 &  -78.3 &  -99.7 & -41.3 & -58.5 & -115.6 & -71.9 \\

\bottomrule
\end{tabular}

\subsubsection*{Absolute binding affinity: evoEF2}

\begin{tabular}{lrrrrrrrrr|r}
\toprule
Method & $t_1$ & $t_2$ & $t_3$ & $t_4$ & $t_5$ & $t_6$ & $t_7$ & $t_8$ & $t_9$ & mean \\
\midrule
            PPO & -49.0 & -35.8 & -40.0 & -47.1 & -49.2 & -47.2 & -24.8 & -37.7 & -53.3 & -42.7 \\
            PPO & -45.8 & -33.4 & -45.1 & -46.0 & -44.9 & -47.8 & -29.3 & -29.0 & -52.1 & -41.5 \\
            PPO & -43.0 & -37.2 & -40.4 & -41.4 & -41.7 & -45.3 & -29.7 & -36.0 & -50.1 & -40.5 \\
                          \hline

           SPPO & -53.3 & -33.4 & -50.6 & -53.9 & -58.8 & -57.5 & -34.9 & -29.5 & -59.4 & -47.9 \\
           SPPO & -49.4 & -37.0 & -50.6 & -53.7 & -56.1 & -56.3 & -34.0 & -35.8 & -59.0 & -48.0 \\
           SPPO & -54.7 & -34.6 & -50.1 & -54.4 & -58.2 & -60.1 & -23.9 & -28.5 & -65.9 & -47.8 \\
                         \hline

 Rosetta design & -57.0 & -38.9 & -51.9 & -52.6 & -54.8 & -53.8 & -38.8 & -38.3 & -59.3 & -49.5 \\
 Rosetta design & -46.0 & -38.1 & -40.4 & -50.0 & -55.8 & -45.6 & -33.1 & -28.3 & -53.8 & -43.5 \\
 Rosetta design & -45.8 & -38.8 & -49.0 & -49.9 & -45.0 & -43.2 & -27.5 & -38.1 & -53.2 & -43.4 \\
               \hline

         Native & -57.0 & -37.0 & -49.8 & -54.6 & -56.5 & -59.3 & -33.3 & -31.0 & -60.6 & -48.8 \\
                       \hline

       Han\&Kral & -45.9 & -39.2 & -41.4 & -39.1 & -52.5 & -57.0 & -36.6 & -34.3 & -52.0 & -44.2 \\
                     \hline

          Huang & -46.7 & -38.6 & -59.7 & -48.8 & -59.4 & -51.4 & -31.7 & -43.4 & -59.7 & -48.8 \\
\bottomrule
\end{tabular}

\clearpage
\newpage
\subsection{Molecular Dynamics}

 \begin{figure}[hb!]
     \centering
     \includegraphics[width=0.3\linewidth]{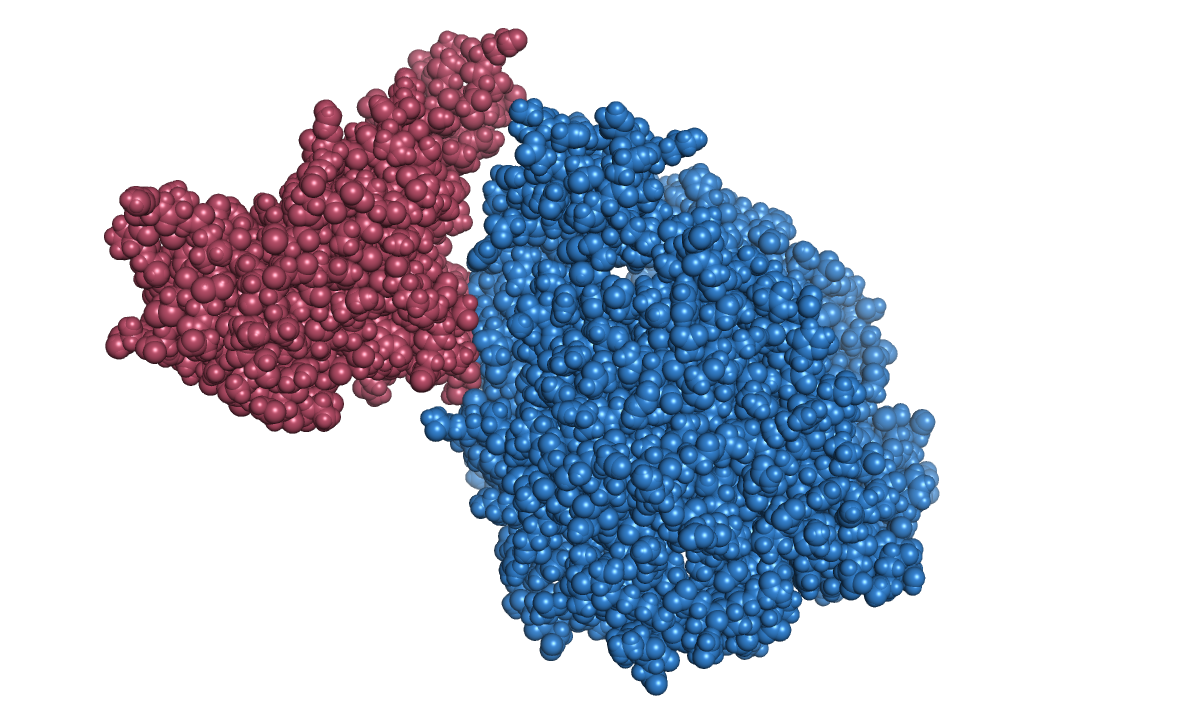}
     \includegraphics[width=0.3\linewidth]{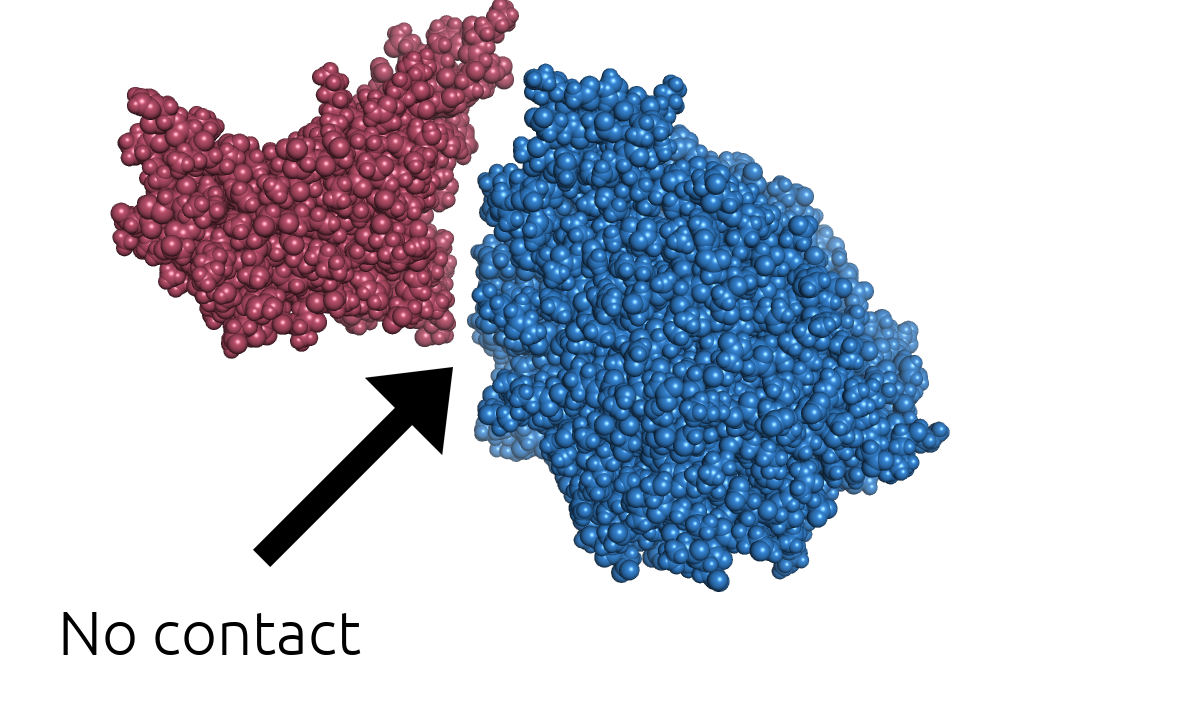}
     \includegraphics[width=0.3\linewidth]{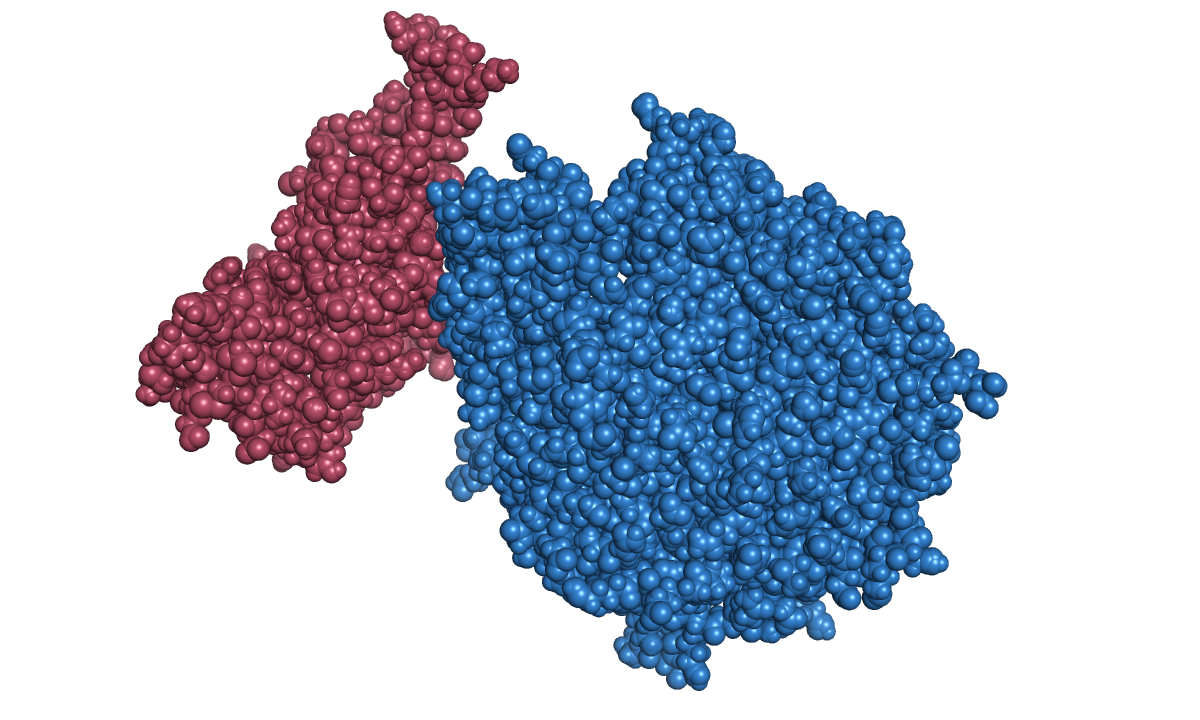}
     \includegraphics[width=0.3\linewidth]{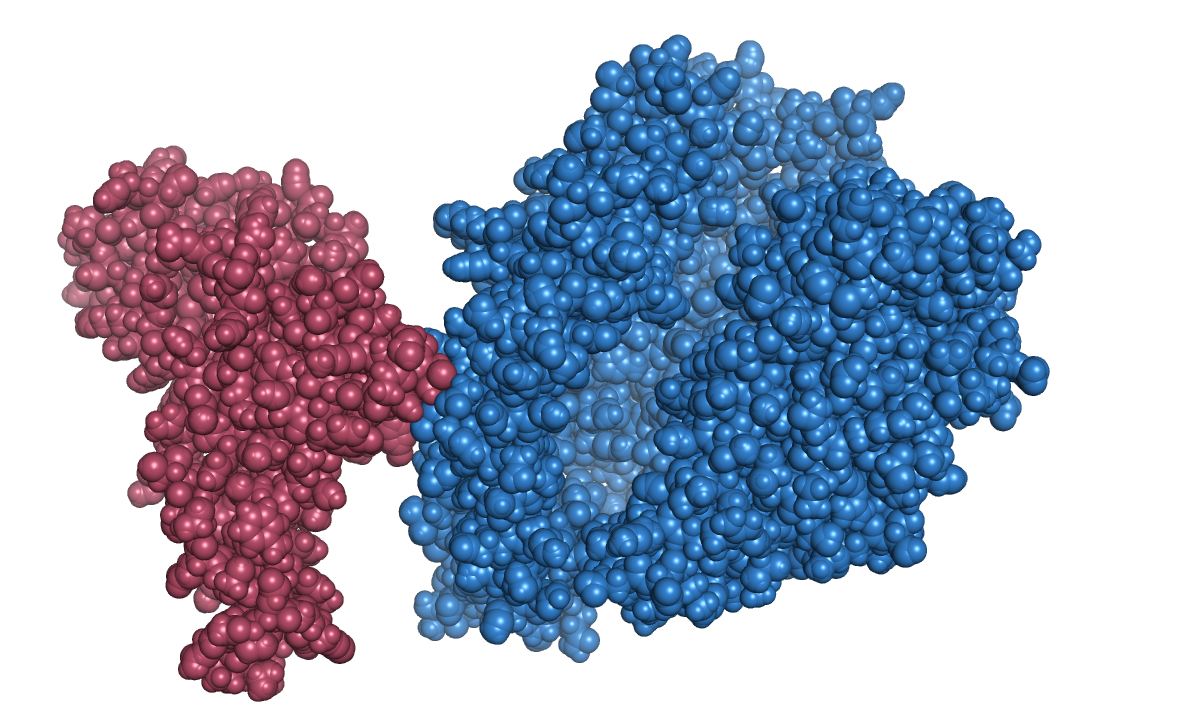}
     \includegraphics[width=0.3\linewidth]{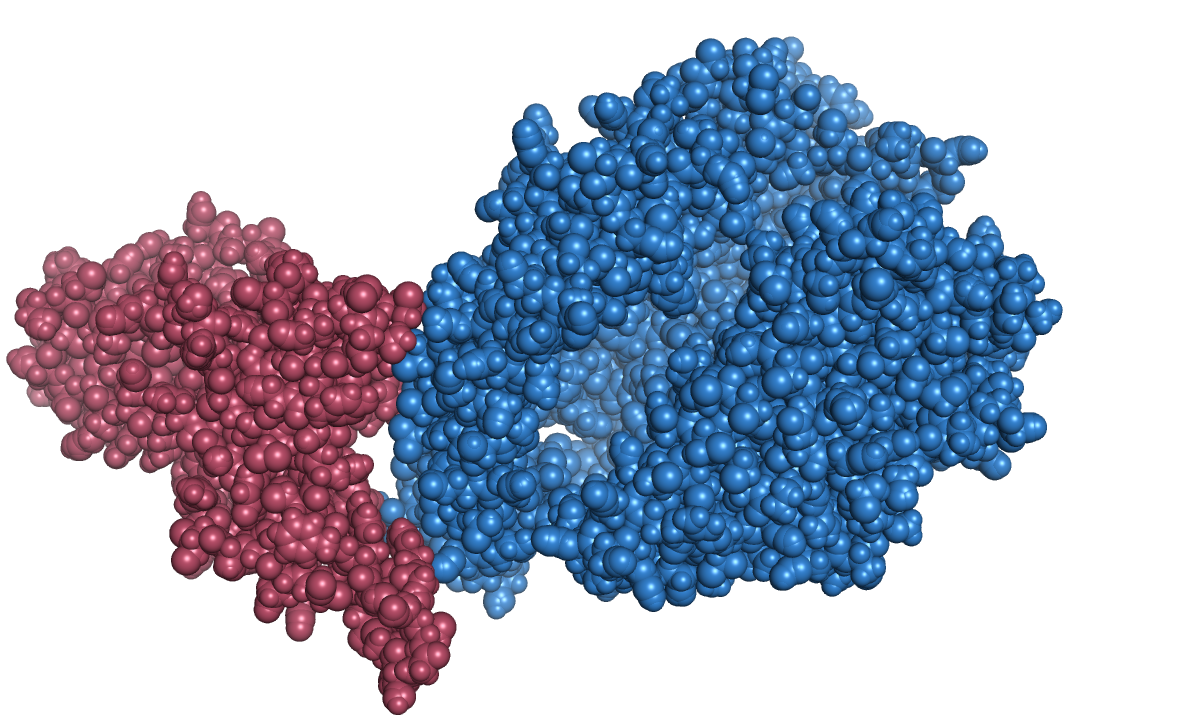}
     \includegraphics[width=0.3\linewidth]{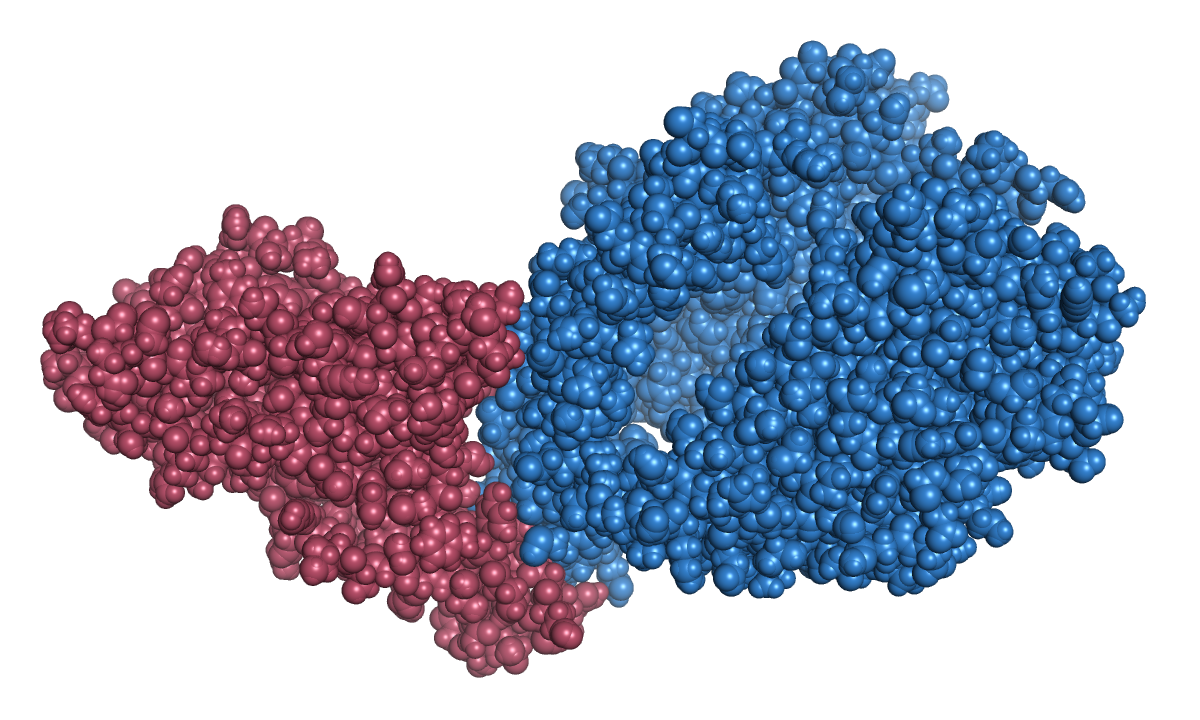}
     \caption{Example molecular dynamics simulation (duration 50ns, with 50 million time steps of 1 fs) trajectory of the \sarscov spike protein (red) complexed with native ACE2 (blue) simulation (top) and one of our top ranked designs (bottom). Left/center/right column: snapshots at 5ns/25/40 ns. Native structure starts from an experimentally determined, bound position, promptly \emph{disassociates} from the protein and intermittently explores binding. In the bottom row, we demonstrate unrestrained binding of the designed protein, which starts nearly unbound, that is is translated away from the complex and rotated by a random small angle, then mimimized as a soluble monomer and translated back to contact with the receptor. The new design starts forming biochemical contacts, and finally forms a \emph{stable complex} with the spike protein. Note that the simulation of the native complex cannot be expected to be typical and is provided here for illustrative purpose only. }
     \label{fig:mdVisualisation}
 \end{figure}
 
 \begin{figure}[thb!]
\centering
\includegraphics[width=.24\linewidth]{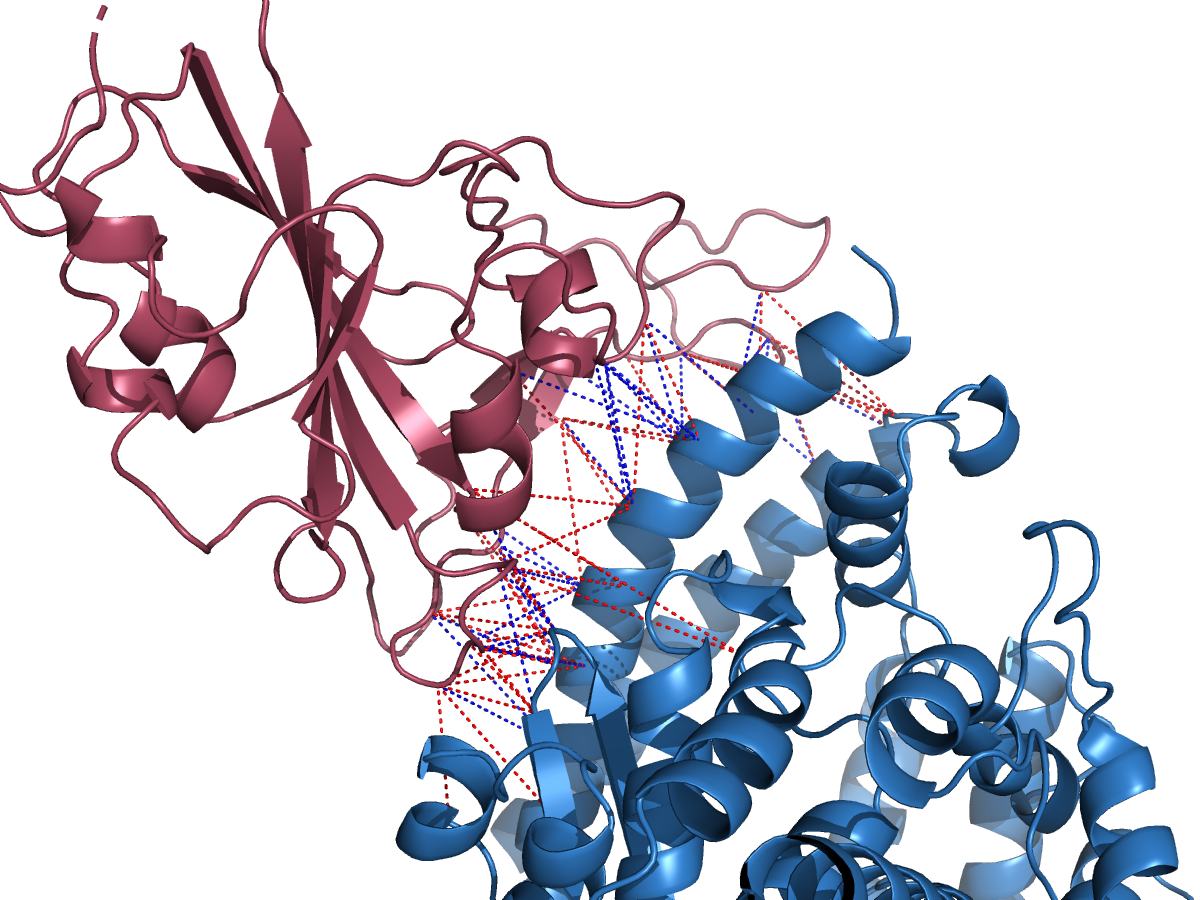}  
\includegraphics[width=.24\linewidth]{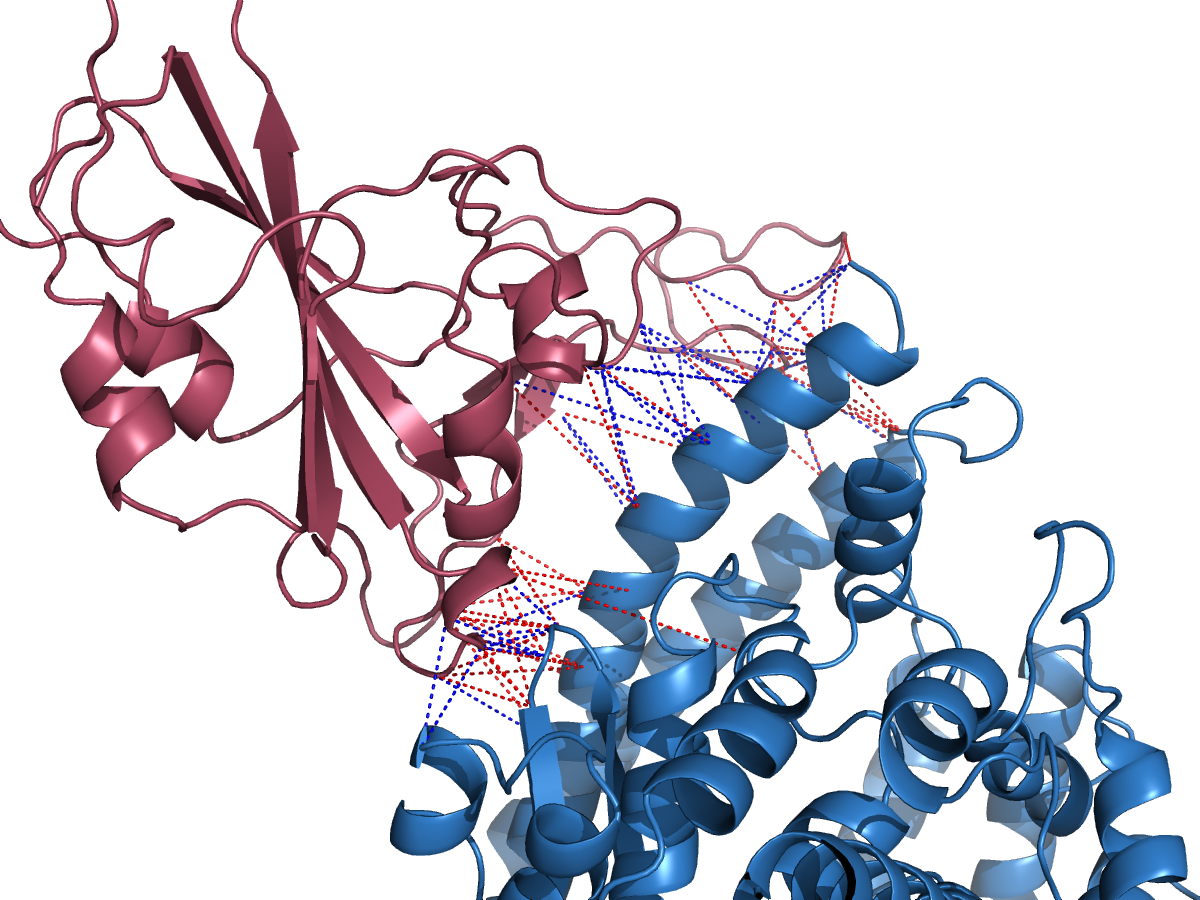}  
\includegraphics[width=.231\linewidth]{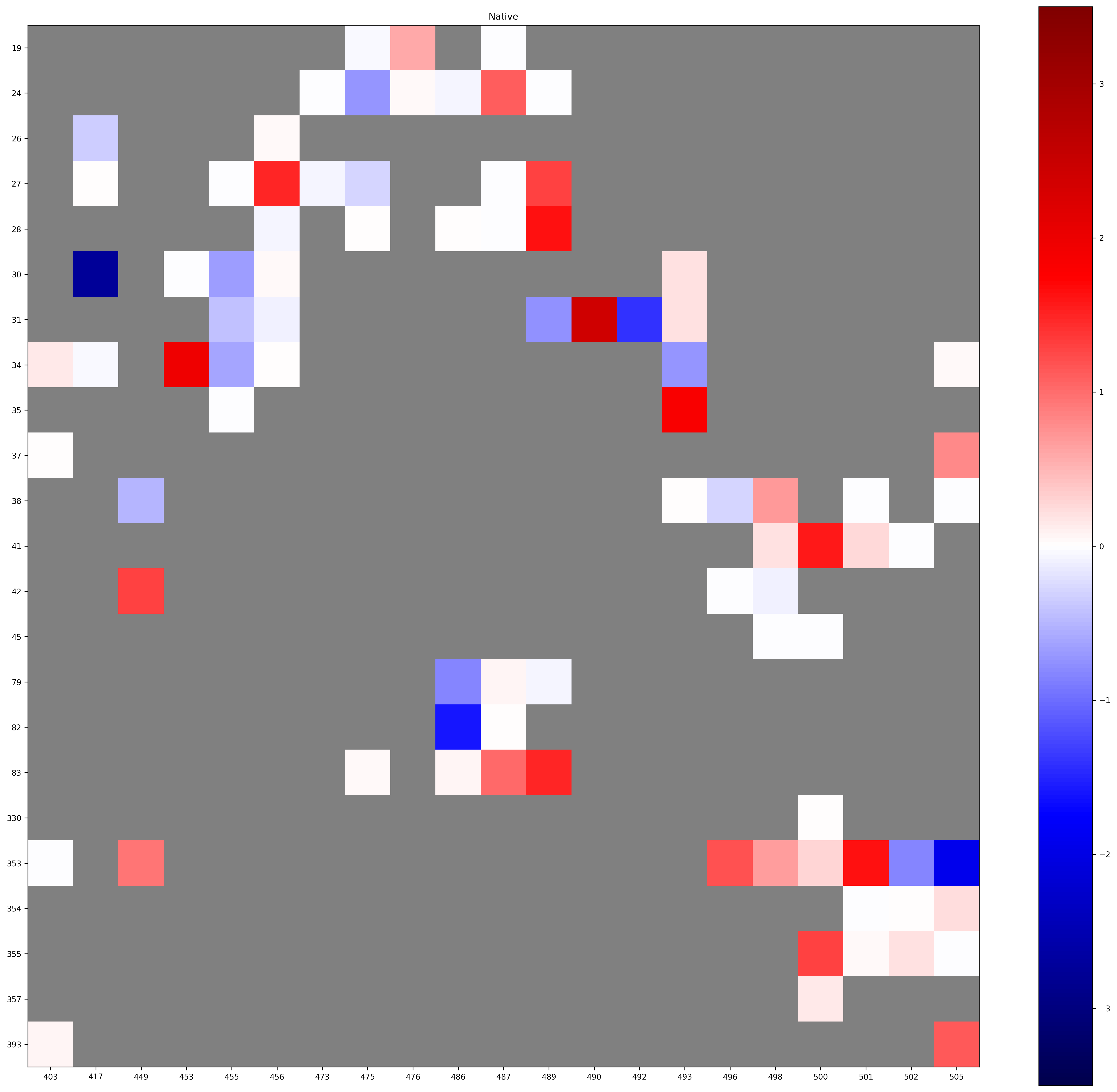}
\includegraphics[width=.2576\linewidth]{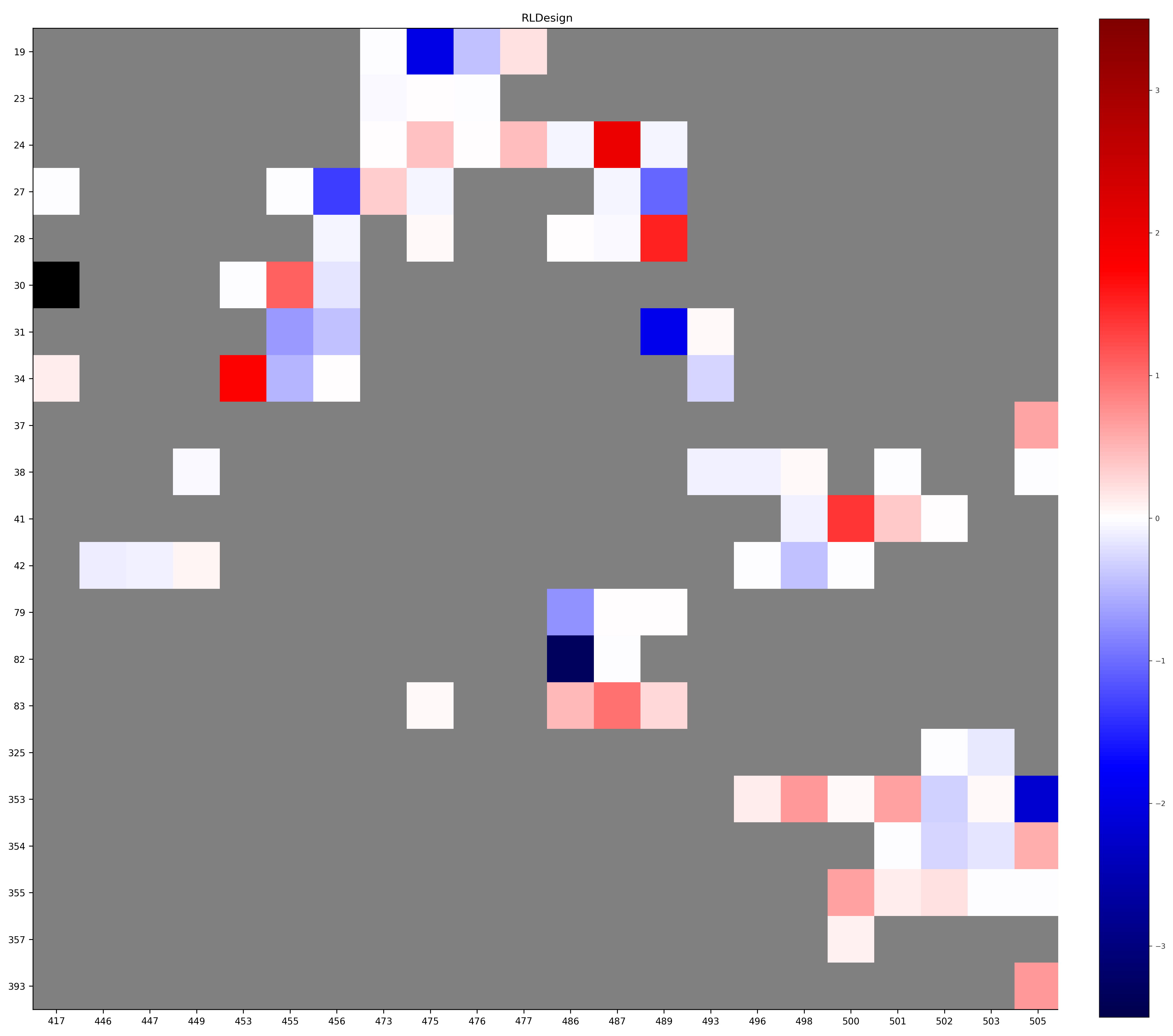}  
\caption{Interactions between chains in the native structure (left figures) and one of the good designs (right figures). Heat maps depict contact maps of all interacting residues, protein structures visualise the 3D interactions. Blue interactions are favorable and have negative energy, red -- unfavorable. Not only does the new design bind tighter, with more residues involved in binding, but also it has fewer detrimental interactions.}
\label{fig:head2head}
\end{figure}
 
For each of the methods, we have selected one design, assessed as the most suitable one by our methodology (that is having the best binding and stability). On each of these structures, we have performed a single run of molecular dynamics simulations. 

These simulations have been performed using GROMACS 2018.4\cite{abraham2015gromacs,pall2014tackling,pronk2013gromacs}, with an atomistic AMBER \texttt{ff03} as a force field. Each of the low-energy poses of a new design based on native ACE2-spike protein co-crystal structure has been fully solvated (with a three point water model - SPC/E) in a cubic box with 20\AA margin. The choice of simulation parameters has not been benchmarked, but each of them is a rational choice, according to the literature. We do not expect that the results of simulation will be radically different, were we to make an other choice.

Each of the solvated systems has been  minimized for the 1 ns (50,000 steps of 2 fs each), equilibrated with respect to temperature and pressure for 1 ns each (with a modified Berendsen thermostat and Parinello-Raman pressure coupling, respectively). Finally we simulated the system unrestrained for 200 ns, in line with the simulations previously reported in literature~\cite{han2020computational}.

\begin{figure*}[ht!]
    \centering
    \includegraphics[width=\textwidth]{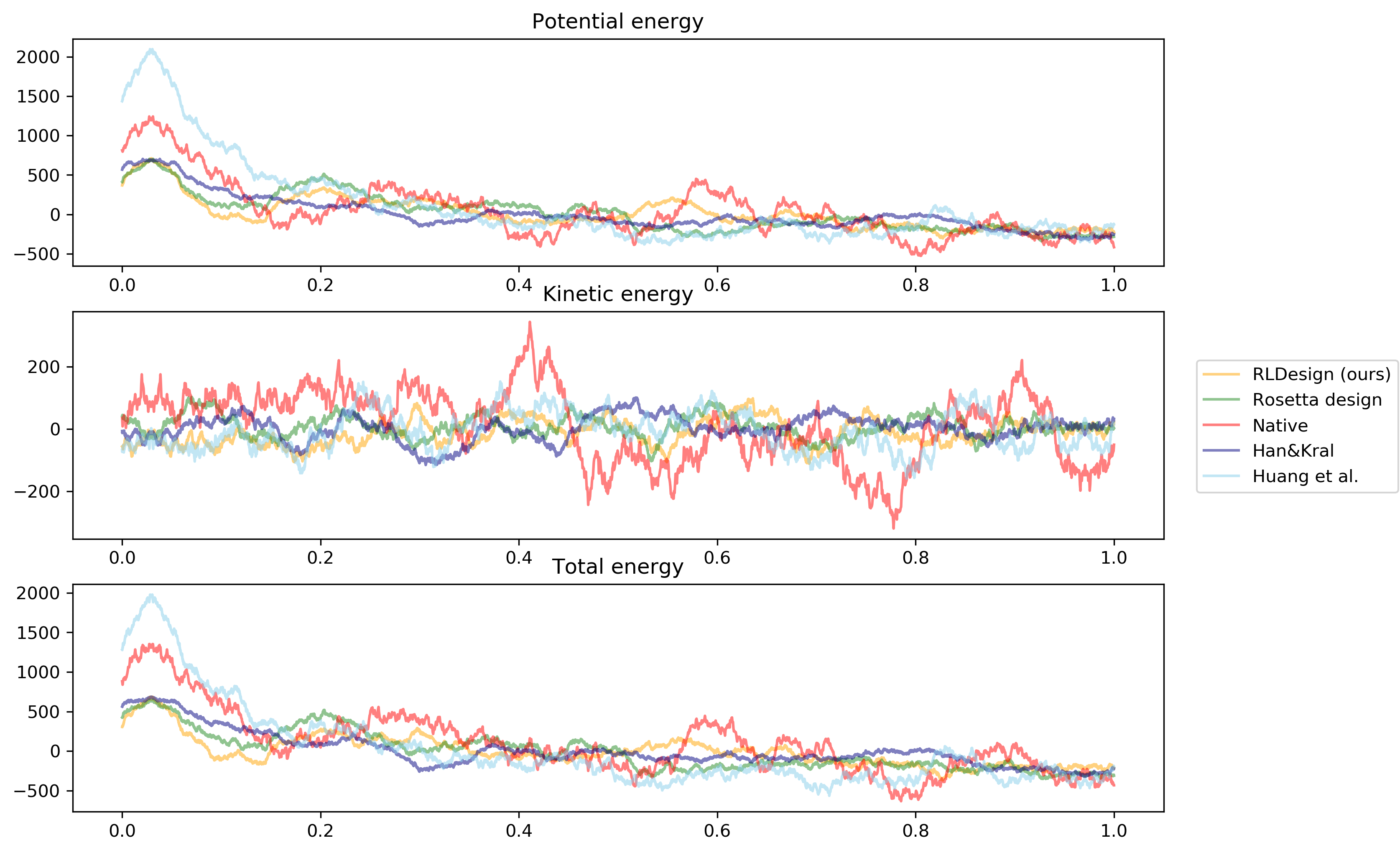}
    \caption{Molecular dynamics. The plots illustrate the development of the energy of a simulated system. As each of the systems comprises different number of atoms, with differently shaped periodic box, we have offset each trajectory by its median value, thus $0$ corresponds to the median value of the observed energy. Large fluctuations represent a frustrated, dynamic system, low -- a system at equilibrium. We note, that -- with exception of the native structure -- every design appears to settle in a comfortable energy well. }
    \label{fig:mdPlots}
\end{figure*}

The simulations reveal (see Figure~\ref{fig:mdPlots}, that each of the trajectories is largely stable, with one of~\cite{huang2020computational} starting in a potentially frustrated state, but promptly settling into an energetically favorable pose. However, we have noticed that the pose based on the native state starts with a large amount of potential energy, which is approximately 40\% into the simulation converted into kinetic energy. This corresponds to the complex disassociating. The final 10\% of simulation, where both potential and kinetic energy are low, illustrates the complex forming anew. 

From molecular dynamics simulations, we can conclude, that each of the designs, be it proposed by our method, or by others, is biophysically feasible and can result in a high-affinity binding. 

\end{document}